\documentclass[a4paper]{article}

\usepackage{amsmath,amssymb,amsthm,graphicx,fixmath,tikz,latexsym,color,xspace}
\usepackage[margin=1in]{geometry}

\usepackage[ruled,noend,linesnumbered]{algorithm2e}
\usepackage{xspace}
\usepackage{wrapfig}
\usepackage{enumerate}
\usepackage{url}
\usepackage{biblatex}
\definecolor{Apricot}{rgb}{0.98, 0.81, 0.69}
\definecolor{Orange}{rgb}{1.0, 0.5, 0.0}

\newtheorem{theorem}{Theorem}[section]
\newtheorem{lemma}[theorem]{Lemma}

\newtheorem{corollary}[theorem]{Corollary}

\begin{document}

\title{The Crew: The Quest for Planet Nine is NP-complete}

\author{Frederick Reiber}
\date{}

\maketitle

\begin{abstract}
In this paper, we study the cooperative card game, The Crew: The Quest for Planet Nine from the viewpoint of algorithmic combinatorial game theory. The Crew: The Quest for Planet Nine, is a game based on traditional trick-taking card games, like bridge or hearts. In The Crew\footnote{During the writing of this paper, another game in the same line, The Crew: Mission Deep Sea, was released. However, when we say The Crew, we are referring to The Crew: The Quest for Planet Nine, and not its sequel.}, players are dealt a hand of cards, with cards being from one of $c$ colors and having a value between 1 to $n$. Players also draft objectives, which correspond to a card in the current game that they must collect in order to win. Players then take turns each playing one card in a trick, with the player who played the highest value card taking the trick and all cards played in it. If all players complete all of their objectives, the players win. The game also forces players to not talk about the cards in their hand and has a number of "Task Tokens" which can modify the rules slightly.

In this work, we introduce and formally define a perfect-information model of this problem, and show that the general unbounded version is computationally intractable. However, we also show that three bounded versions of this decision problem - deciding whether or not all players can complete their objectives - can be solved in polynomial time.
\end{abstract}


\section{Introduction}
For over 100 years, mathematicians have been using mathematics to analyze games, with one of the originals being the game of Nim \cite{nim} . Since then, an entire field of mathematics, combinatorial game theory, has been created to analyze games. More recently, researchers have started using computation as a means to analyze games, which has also lead to a number of new ways of analyzing games. One of the most popular questions is discovering the computational complexity of a game, or how hard it is to decide the winner or winners of a given game. This field has started to be called algorithmic combinatorial game theory \cite{CAGT}, in order to separate it from the classical economical focused algorithmic game theory \cite{AGT}. A number of classic games have been analyzed through the algorithmic combinatorial game theory lens, including Hex \cite{hex}, Go \cite{go}, and Chess \cite{chess}. Additionally, some more recently results have shown the complexity for more 'hobbyist' games including Pandemic \cite{pandemic}, Kingdomino \cite{kingdomino}, and Hanabi \cite{hanabi}.

In this work, we study the computational complexity of the cooperative trick-taking card game, called The Crew: The Quest for Planet Nine. Designed by Thomas Sing and published in 2019 \cite{bggCrew}, the game has received a number of accolades, including the prestigious 2020 Kennerspiel des Jahres \cite{award}. In the game, players play as astronauts working together to discover the mysterious ninth planet located at the edge of the our solar system.

\subsection{Rules of the Game}
The Crew: The Quest for Planet Nine, is a cooperative trick-taking card game for 2-5 players, in which everyone must work together to complete all the objectives. The game features a campaign system in which groups of players work through the 50 increasingly difficult missions. The game's main component is 40 card deck. Each card has two attributes. The first is a value that ranges from 1 to 9 for the non trump suits, and 1 through 4 for the trump suit. Each card is also assigned a suit(color): pink, yellow, green, blue, and rocket(black) with rocket being the trump suit. The game also contains a second deck of cards that represent the objectives. Each of these maps to one of the non trump cards.

To set up the game, players deal out all the cards to the players, as evenly as possible (because 40 is not a multiple of 3, when playing with three players, one player will receive an extra card). Then, players draw a number of objective cards, as determined by the scenario they are playing, and apply and scenario specific rules including, applying one of the task tokens. Task tokens can be thought of as modifiers to specific objectives, adding another requirement to complete said objective. Generally, this requirement deals with the order in which objectives are completed. Players then draft the drawn objective cards, determining who must complete which objective.

Once set up has finished, play begins. Players taking turns playing a single card face up in the middle of the table. This is referred to as the "trick". The first card played, sets the suit of trick. All other players must play a card with the same suit if able. Otherwise, they can play any card in their hand. Once everyone has played a card, the winner is the player with the highest value card that matches the suit of the trick. However, if another player has played a trump card, the highest trump card takes the trick, although other players must follow the original suit if possible. That player then collects all cards played in the trick, and plays the first card of the next trick. Once a card has been played in a trick, it can not be played again. If any of the collected cards are on objectives owned by the trick's winner, that player marks them as completed. If all objective cards are completed in such a way that all task token requirements are satisfied, all players win. However, if a single objective is failed all players lose. There are two ways to fail an objective. The first is for a different player to collect the card described on the objective. The second is for the objective to be completed out of the order specified by the task tokens. The final way for the game to end is for any number of players to run out of cards. In this scenario, all players lose.

Finally, the game is also an imperfect information game, with players only able to know what cards are in their hand. As players are working together, the rules forbid players from communicating about the cards in their hand. The game has a system for limited communication, but as this is not critical to our analysis, we will not discuss it here. For the full rules of The Crew, see Appendix A.

\subsection{Related Work}
As mentioned in the introduction, a number of games have had their computational complexity analyzed. Both Kingdomino \cite{kingdomino} and a generalized version of Pandemic \cite{pandemic} have been show to be NP-complete. One of the common themes in most of these analyses, is that the games parameters must be unbounded. Otherwise, most algorithmic questions would technically be solvable in constant time through a simple brute force technique. As such, we will define a unbounded version of this game that will be used in our analysis.

There are also a number of card games that have been shown to be NP-complete. Some of these include, Hanabi \cite{hanabi}, Uno \cite{uno}, and FreeCell \cite{freecell}. When playing these games, as well as The Crew, there is usually some amount of unknown information; however, this makes analysis significantly more difficult. Thus we will be using a perfect information model, something common in card game analysis. Another common simplification made in card game NP-complete proofs is the reduction to a single player. While Uno and Hanabi are traditionally multiplayer games, both reductions given rely on a single player version of the problem. However, due to the nature of The Crew, doing so would produce a game that is trivial, as a single player would control all cards. Instead we will treat the number of players as a parameter of the problem. We will also make the assumption that all players act in accordance with what is in their best interest.

There are also a number of non NP-complete complexity results. Many popular games have had complexity results found. Some of these include Hex \cite{hex}, Chess \cite{chess}, and Go \cite{go}. One surprising result is that the popular card game, Magic: The Gathering has been shown to be Turing complete \cite{magic}. 

Finally, general trick-taking card games, which are traditionally non-cooperative, when looked at as a perfect information problem, have shown to be P-space complete \cite{ttcg}. There is also a number of works when dealing with trick-taking card games as a artificial intelligence problem \cite{bridge1} \cite{bridge2} \cite{bridge3}. However, at the time of writing, there is no known research on cooperative trick-taking card games.

\subsection{Results and Organization}

In this work, we look at the algorithmic and computation complexity of the The Crew: The Quest For Planet Nine in terms of of the number of cards, both value and suit, the number of players, and the number of objectives. In Section 2, we fully define the model and the parameters for this problem. In Section 6 we show that the problem is NP-complete when all parameters are unbounded.

With this result, we instead look at designing algorithms for particular cases of the problem. The first of the three cases is one in which all cards have different suits. For this instance we give a simple combinatorial algorithm. The second is the instance in which all objective cards (cards that are referred to by objectives) are owned by the player who possesses said objective and all cards share the same suit. Note that this second requirement is important; otherwise, as we later show, the problem in NP-Hard. For this instance of the problem, we provided a greedy algorithm that is almost linear in terms of $n$. Finally, we look at version of the problem with the only restriction being that all cards have the same suit. This algorithm runs in polynomial time in terms of all parameters. The following table gives exact running times for all algorithms introduced in this paper.

\begin{table}
    \centering
    \begin{tabular}{|c|c|c|}
        \hline Instance of Problem & Running Time & Section.\\ \hline
        Single Value & $\mathcal{O}(n^2)$ & Section 3 \\
        Single Suit, Owned Objectives &  $\mathcal{O}(np + n\log{}n)$ & Section 4\\
        Single Suit &  $\mathcal{O}(n^2 + p^2 + np\log n)$ & Section 5 \\
        General Case & NP-Complete & Section 6 \\ 
        General Case with Task Tokens & NP-Complete & Section 6. \\
        \hline
    \end{tabular}
    \caption{Summary of results where $n$, $p$, and are number of cards and number of players respectively.}
    \label{tab:my_label}
\end{table}

\section{Definitions and Model}

All cards in The Crew, have two attributes, a \textit{value} and a \textit{suit}. We define a \textit{card} $c$ to be a tuple $(x, y)$ where $x \in \{1, ...,k\}$ is the cards value, and $y \in \{1,...,s\}$ is the cards suit. Unlike in an actual game of The Crew, our model does differentiate between trump and non trump suits. This is done to simplify analysis, but will also be touched on later in this work. A game of The Crew, also has a finite number of $p$ players, where $p \geq 1$. As mentioned above, our analysis will be focusing on when $p \geq 2$, as when $p=1$ the game is trivial. During the set up of The Crew, each card from from set $C$ with size $n$, where $n \leq k*s$, is dealt to an individual player $i$. Each player's set of cards, is often referred to as their "hand" and will be notated as $C_i$. It is important to note that we make no assumption that each hand is the same size. If a player's hand is ever empty, the game is over. Additionally, The Crew's deck of cards only has 1 card per value and suit combination. As such, we keep $C$ to the standard definition of a set, meaning each card in our problem instance is unique.  Finally, we use the following notation to show that a specific card is in a player's hand: $c_i$ or $(x,y)_i$.

We define playing a trick $t$ to be selecting one card from each hand $C_i$ starting from the lead $\tau \in [1, p]$. Formally a trick is $\{c_\tau \in C_\tau, c_{\tau+1} \in C_{\tau+1}, ..., c_p \in C_p, c_1 \in C_1, ... c_{\tau - 1} \in C_{\tau-1}\}$. As per the rules of the game, it is required that suits are followed, or that each card played matches suits with the first card played by player $\tau$, as long as the corresponding hand $C_i$ has a card in that suit. Formally this is: $y_{c_i} = y_{c_\tau} \lor \forall c \in C_i, y_c \neq y_{c_\tau}$. Once a card has been played in a trick, it can not be played in any further tricks. The winner of a given trick, is the card with the highest value and suit equal to the suit of the lead card. The hand which then wins the trick is now the lead and, assuming play continues, will lead the next trick. During the first trick, when no lead $\tau$ has been determined, we allow any player to be the lead and leave that decision as part of the problem statement.

The last component we need define is the objective. Since we are not using trump cards in our analysis, we can use the following definition. An objective $o$ is a tuple $(c, z)$, where $c$ is the card required to complete it (also refered to as an "objective card"), and $z$ is a boolean that represents whether the objective has been completed or not. In order to complete an objective, the player assigned to said objective must collect a trick in which card $c$ has been played. We use $O$ to signify the set of all objectives that are in the game, and define its size to be $l$. To signify that an objective has been assigned to a specific player we write: $o_i$, and use $O_i$ for the set of all objectives assigned to a single player. We also put the following restriction on all objectives: $\forall o \in O, \, \nexists q \in O, c_o = c_q$. Written out, this means that no two objectives can have the same target card.  

With all important components defined, we can now formally define the problem statement. We define a instance of The Crew, to the following decision problem. Given a distribution of the set of cards $C$, as well as a distribution of the objective set $O$, determine whether or not a sequence of tricks exists that will complete all objectives in $O$. We use the following parameters to bound the problem, $k$, $s$, $p$, and $O$. We use the following shorthand to denote various versions of the problem $\beta(k,s,p,O)$, with $\beta(\_,\_,\_,\_)$ being the largest class of problems in which no bounds are put on the problem statement.

\begin{figure}

\centering

\begin{tikzpicture}

\node[align=center] (p1) at (0,0) {$p_1 - \tau$\\$o_1=\{((3,1),f)\}$\\$s_1:2,4 $\\$s_2:1,3$\\$s_3:-$};
\node[align=center] (p2) at (5,2.5) {$p_2$\\$o_2=\{((4,2),f),((7,3),t)\}$\\$s_1:3$\\$s_2:4$\\$s_3:1,2,$};
\node[align=center] (p3) at (10,0) {$p_3$\\$o_3=\{(7,3),t\}$\\$s_1:- $\\$s_2:5$\\$s_3:3,4$};
\node[align=center] (p4) at (5,-2.5) {$p_4$\\$o_4=\{\}$\\$s_1:1 $\\$s_2:2$\\$s_3:5,6,7$};

\path[->] (p1) edge [bend left, thick] (p2)
    (p2) edge [bend left, thick] (p3)
    (p3) edge [bend left, thick] (p4)
    (p4) edge [bend left, thick] (p1);

\end{tikzpicture}

    \caption{An instance of The Crew: The Quest for Plant Nine with a valid play sequence. To complete both remaining objectives, $p_1$ should lead with $(4,1)$, and $p_3$ should "sluff" $(5,2)$. For the second trick, $p_1$ should lead with either $(1,2)$ or $(3,1)$.  }
    \label{fig:my_label}
\end{figure}
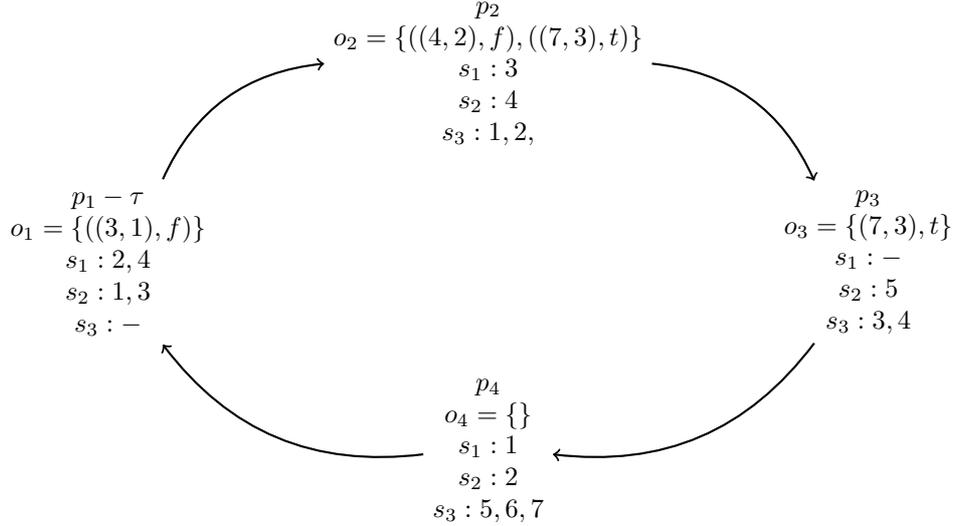

\section{Single Value}

To help introduce this problem, we will first consider the (some-what silly) instance of the problem in which the maximum value for a card is 1, or the $\beta(1,\_,\_,\_)$ instance version of The Crew. 

\begin{theorem}
The $\beta(1,\_,\_,\_)$ instance of The Crew can be decided in $\mathcal{O}(n^2) time.$
\end{theorem}

Before giving the algorithm, we will first discuss the two observations critical to it. Notice that by limiting the value to 1, we are also limiting the number of cards in a given suit to 1. Recall that the winner of a trick is the hand that played the highest value of the led suit. However, because no cards can share a suit, whichever hand leads will win the trick. By extension, which ever player starts with the lead, will be the only one who can take tricks. Thus, if more than a single player possesses any objectives ($O\neq O_i$), no successful play sequence can exist. This also means that which ever player starts in the lead, can take all cards at all times, as players are under no requirement to follow the lead suit. Therefore, the only time in which a single player $i$ controls all objectives ($O=O_i$), does not lead to a valid play sequence is one in which a player runs out of cards before all objectives can be completed. 

To decide the $\beta(1,\_,\_,\_)$ instance of The Crew, we first iterate through all objectives and check that only a single player possesses objectives. If this is not the case, return false. If this instance has no objectives, return true. Then, iterate through all players, and find the minimum hand size. If during this process a hand has 0 cards, return false. Next, we iterate through all objectives, mapping them to the player who possesses the objective card and keeping a counter of how many objective cards each player has. Finally, we compare the maximum number of objectives cards a single player has to the minimum number cards a single player has. If the difference between these two values is $\leq 0$, then a valid play sequence exists. If the difference is greater, no such play sequence exists. 

In terms of run time, iterating through the objectives takes at worst $\mathcal{O}(l)$ and iterating through all the players takes $\mathcal{O}(p)$ time. The last step, assuming set operations are constant, takes $\mathcal{O}(lp)$ time. However, because each objective must map to a unique card we can say that $l\leq n$. Additionally, we can also bound $p$ in terms of $n$. Notice that if we ever reach a hand that has zero cards, we return. If we assume that $p > n$, at some point during the second step, we must return, as not all players will have a card. At worst, this requires $n+1$ iterations, thus we can bound $p$ with $n$. This gives a total worst case run time of $\mathcal{O}(n^2)$.

\section{Single Suit and Owned Objective Cards}
In this section we look at a version of our decision problem where every card in the deck is of a single suit, or were $s = 1$, and all objective cards are owned by the player with that objective. Written mathematically: $\forall o \in O,\, c_{o_i} \in C_i$. We define this set to $O^*$, and define this instance of our game to be $\beta(\_,1,\_,O^*)$.  We give a polynomial time algorithm to decide this version of the problem, using a greedy heuristic. Note that we make no assumption on any of the other parameters, although as defined earlier we do not allow duplicate cards in our model. 

\begin{theorem}
\label{ssoo:algo}
We can decide the $\beta(\_,1,\_,O^*)$ instance of The Crew problem in $\mathcal{O}(np + n\log{}n)$. time.
\end{theorem}

To decide $\beta(\_,1,\_,O^*)$, we introduce the following algorithm. To start, select an objective from the set of objectives ($o \in O$), that has not been completed ($z_o = false)$. Then, find which player ($i$) is assigned that objective and have that player play that objective card $c_o$, in the trick. Then, have each following player play the card with the highest value that is less than the already played objective card, and is not an objective card. This can also be written mathematically as $max\{c \in C_i : x_c < x_{c_o} \land c \notin O_c \}$. If no card exists that satisfies this condition we can return false. Once this trick has concluded, repeat the process with a new objective until all objectives are completed or one objective returns false. To prove that algorithm is correct, we first present the following two lemma.    

\begin{lemma}
\label{ssoo:tricks}
For the decision problem $\beta(\_,1,\_,O^*)$ if a successful play sequence exists, there must be a play sequence that requires $l$ tricks.  
\end{lemma}

We prove this lemma in two parts. First, notice that all objective cards are owned by the players who must complete said objectives. Because each player is limited to playing only $1$ card per turn, this means only one objective can be completed per trick. As such, a successful play sequence must contain at least $l$ tricks. Second, consider the instance in which a greater than $l$ trick play sequence exists. Because all cards can be played for their value at all times, we can construct $l$ trick sequence from a $l+1$ trick sequence by simply removing the trick which does not complete an objective. This does not affect the success of the sequence, as the following trick can still be constructed to have the same outcome by all players playing the same card. Thus, any successful sequence with greater than $l$ tricks implies the existence of a successful sequence of $l$ tricks.

\begin{lemma}
\label{ssoo:cards}
If playing the "best" card ($max\{c \in C_i : x_c < x_{c_o} \land c \notin O_c \}$) when attempting to complete one objective, in the $\beta(\_,1,\_,O^*)$ problem, would make another objective impossible to complete, it is impossible for both objectives to be completed.
\end{lemma}

To start, look at what must be required in order for an objective to be completed. Recall that all objective cards are owned by the player who possess its corresponding objective, so in order to complete an objective, one must be able to take a trick with the objective card. This also means that only objective cards should be winning tricks. Because no cards can be "sluffed", meaning to play a card without it contributing to deciding who wins the trick, each hand must play a card with a value lower than the objective card. For the sake of contradiction, assume that some other play would allow for both objectives to be completed. We will call this card $b$, and our original play card $a$. Because card $a$ is also lower than the objective card, switching it with $b$ in this trick does not change the outcome. Additionally, because $b$ has, by definition, a lower value than $a$, card $b$ can be played in any trick in which $a$ was to be played. This contradicts our earlier assumption, that playing $a$ resulted in another objective being impossible to complete, thus we know the lemma to be true.  

With both lemma proven, it is easy to see why our greedy algorithm is correct. As per lemma 4.2, it is sufficient to only search for play sequences that require $l$ tricks, and as per lemma 4.3, our algorithm will never make a bad choice (assuming a good choice exists). Thus, our algorithm properly decides the decision problem $\beta(\_,1,\_,O^*)$.

In terms of run time, we can show that each objective requires almost linear time. The first major question is how to answer the "best" card queries, which are essentially   predecessor queries. Traditionally we would use a balanced binary-search tree, which can answer predecessor queries in $\mathcal{O}(\log{}n)$ time. However, since all elements can be sorted into a singe list, we can use Dynamic Fractional Cascading \cite{dynamic} which can answer all queries for an individual objective in $\mathcal{O}(p\log{\log{n}} + \log{}n)$ time.  Luckily, due to only needing to remove cards from the structure, the $\mathcal{O}(\log{\log{n}})$ factor is reduced to $\mathcal{O}(1)$. We will make a total of $l$ full queries. Additionally, we can handle delete operations in $\mathcal{O}(\log{}n)$ time and for worst case $n$ delete operations in $\mathcal{O}(n log{}n)$ time. Set up of the structure requires $\mathcal{O}(n)$ time, as does separating all objective cards. Finally, to avoid removing/playing a card from the hand who is playing the objective card, we create a dictionary and check it before every deletion operation. Checking takes $\mathcal{O}(1)$ time, and setting it up takes $\mathcal{O}(n)$ time. Because $l \leq n$ as each objective maps to a unique card, we can say that total time for the algorithm is $\mathcal{O}(np + n\log{}n)$.

\section{Single Suit}

We will now look at a version of the problem with decreased bounds from our earlier version, specifically the requirement that all objective cards are owned by the player ($O^*$). Formally, this version of the problem is written as $\beta(\_,1\_,\_)$ As mentioned in section 1.3, this version of the problem is tractable, and we present a polynomial time algorithm to decided it.

Before discussing our algorithm, we will prove the two lemmas critical in proving its correctness. First, see that the second half of the argument in lemma 3.2 still applies here. Thus if a valid play sequence exists, one must also exists with $l$ or fewer tricks. However, because we have removed the $O^*$ bound, multiple objectives can now be completed in one trick. We will now show that is in fact optimal to do so.

\begin{lemma}
In the $\beta(\_,1\_,\_)$ instance of The Crew, it is optimal to complete as many objectives as possible in a single trick.
\end{lemma}

For the sake of contradiction, assume that an optimal sequence $\sigma$ exists in which we take more tricks than necessary to complete a set of objectives that could be completed in a single trick. From this, we can construct a new single trick $\mu$, that leaves us with more cards. Because all cards can be played at all times, we can always construct the tricks that occur after $\sigma$. Moreover, the set of possible states after $\sigma$ will always be a subset of all possible states after $\mu$.

To construct $\mu$ we use the following rule. If an objective card was played in $\sigma$ play it. Otherwise, play the highest value card out of all that you played in $\sigma$. Because all tricks were taken by the same player, $i$, the highest card in $\mu$ will still be owned by player $i$. With this construction, we have now shown the lemma to be true.\\

Before presenting and proving the second lemma, we will define some new terminology. We will use the term critical player when one of the following two conditions is satisfied. First, the player is attempting to complete one or more objectives this trick. Two, the player is playing an objective card for another player to take this trick. Note that the number of critical players is not always 2. There can be only 1 critical player if that player has the objective card for the objective being completed. There can also be more than 2 players if more than one objective is being completed in a single trick. Obviously, objective cards should only be played when a player is a critical player.  With this definition we can now state the second lemma for this problem.

\begin{lemma}
When dealing with the $\beta(\_,1\_,\_)$ instance of The Crew, the optimal card to play when not a critical player, is the non-objective card with the highest value, that will not win the trick, as long as that card is not one of the $j$th highest cards, where $j$ is the largest number of objective cards owned by a different single player that map to the original player's objective.
\end{lemma}

First, observe that if a player $i$ has any objectives, they will need to, at some point in the sequence, take some number of tricks. Because of lemma 5.1, we know that it is optimal to complete as many objectives as possible in a single trick. Thus, when playing cards we must save the $j$th highest cards, where $j$ is the number of tricks needed to complete all of $i$'s objectives, in which $i$ does not possess the corresponding objective card. We will call this set $a$. For the sake of a contradiction, assume that another set of cards $b$ was instead optimal for completing objectives. By definition, set $b$ must include a card that is not one of the $j$th highest cards, and due to lemma 5.1, we know that both sets have the same cardinality. However, we can create a more optimal set by swapping out all cards not in $a$, for cards in $a$. By definition, the cards only in $a$, must be larger in value, so the outcome of the trick is not changed. We are now left with more cards of lower value, which will then not effect the non critical player tricks. Thus, we have our contradiction, proving that is optimal for a player to save their $j$th highest cards.

To prove the second part of the lemma, we use an argument that is similar to the argument for lemma 3.3. Assume that a more optimal choice for what card to play exists. We will call our original play $a$, and the more optimal play $b$. Because we know that we will only be taking tricks with the $j$th highest cards or objective cards, we can assume the that a more optimal choice is one in which it allows more tricks to not be taken by us. We can infer that $x_a > x_b$, as no cards can share a value, and if it has a higher value, than it falls into one or both of the following cases. Either $b$ will take the trick, per our definition, which will result in a failed play sequence. Or, $b$ is one of the $j$th highest cards, and we know already it is optimal to save such cards. This inference exposes the contradiction. By definition $a$ must be small enough to not win the trick, so switching $a$ and $b$ does not cause any issues in that trick. In later tricks, any instance where $a$ can be used $b$ can also be used as we know $x_b$ is smaller than $x_a$. Therefore, playing $b$ is not any more optimal than playing $a$.  

With both lemma proved, we can now discuss the full algorithm to decide $\beta(\_,1\_,\_)$. 

\begin{theorem}
The $\beta(\_,1\_,\_)$ instance of The Crew can be decided in $\mathcal{O}(n^2 + p^2 + np\log n)$ time.
\end{theorem}
 
To start, find the objective with the largest mapped card, and add that card to the first trick. We will then have player $i$ add their highest card, to the trick. In the case where those two cards are the same, we will play the single card. Then all players who have objective cards that map to player $i$'s objectives, add them to the trick, starting from the objective cards with the highest value. Finally, any players who have not put a card in the trick play according the lemma 5.2. This process repeats until either an objective card is won by the incorrect hand, a player runs out of cards, or all objectives have been completed. 

We will now prove the correctness of the algorithm. With lemma 5.1 and 5.2 proved, we still have two decisions left to prove. The first is that starting from the largest value objective will find a valid sequence if one exists. The second is that critical players play optimally with our algorithm.

To show that it is correct to start with the largest objective, assume that some other play sequence exists ($\sigma$) that does not start with the largest objective. We will now show that we can always construct a valid play sequence that does start with the largest objective, thus if one exists, starting from the highest value objective will properly decide the problem. To do so, simply reorder all the tricks in $\sigma$, so that the objectives are completed in descending order. If one trick completes more than one objective, use the highest value objective for ranking. Because all cards can be played at all times, this does not change the outcome, nor does it impact the tricks construction.       

We will now show that non-objective critical players also play optimally with this algorithm. Assume for the sake of contradiction, that some other card $b$ is optimal when taking a trick other than the current highest card $a$. We will now show that switching the cards, still produces a valid play sequence. There are two cases to consider: The case where we take no more tricks, and the case where we do take more tricks. For the first case, we can switch $a$ and $b$ and still have an valid sequence. Because $a$ has a higher value, $a$ can take a trick anytime $b$ can. Additionally, because $b$ has a lower value, $b$ can be played as a non critical player anytime $a$ can. The same can be said for the second case. With $x_a > x_b$, any time $b$ can be used to take a trick, so can $a$. Recall that we are currently completing the highest objective. If $b$ can complete the current objective, than $b$ can also complete any later objectives, thus switching the two cards still produces a valid play sequence. We have now shown that non-objective critical players play optimally.
 
With both of these two decisions now shown to be correct, we can see that the given algorithm does properly decided the $\beta(\_,1\_,\_)$ of The Crew. We will now discuss the runtime of the algorithm. First, we must create a sorted list of all objectives. Doing so takes $\mathcal{O}(l\log l )$ time. The next step is to calculate $j$ for all hands. For each player, we create a dictionary that maps each other player to a counter and iterate through each of the original player's objectives counting the number of mapped objective cards per player. Once finished we find the max. Doing so takes $\mathcal{O}(l + p^2)$ time, as we iterate through all objectives at most once, but for each player we must iterate through all other players. Once $j$ is calculated we sort each player's hand, taking $\mathcal{O}(pn\log n)$ time. We can then separate the $j$ highest cards from each hand, in constant time per player. The main play loop goes through at most $l$ iterations. During each iteration we first find the current max value objective, which takes constant time. Finding the winning card also takes constant time. We then iterate through all of the winning player's other objectives, and play as many as possible. At worst, this is $l$ iterations. Finally, we answer the predecessor queries, again using Dynamic Fraction Cascading \cite{dynamic} as discussed in section 3, with the cost per trick being $\mathcal{O}(p + \log n)$. Thus the total time for the entire play loop is $\mathcal{O}(l^2 + lp + l\log n)$, with a total time for the algorithm being $\mathcal{O}(p^2 + l^2 + lp +  l\log{l} + pn\log n + l\log{n})$. Again, as $l \leq n$, we could also substitute $n$ for $l$ and get a more generalized upper bound, $\mathcal{O}(n^2 + p^2 + np\log n)$.

\section{NP-Completeness of The Crew}

In this section we provide a proof that \textbf{$\beta(\_,\_,\_,\_)$} is NP-Complete, or when all parameters of the problem statement are unbounded. We also give two modifications to the reduction to show that The Crew is still NP-Complete with more complicated mechanics. Said mechanics are taken from the games rules, which can be viewed in appendix 1. 

\begin{theorem}
The Crew is NP-Complete when all parameters of the problem $(k, s, p, O)$ are fully unbounded, thus giving us the $\beta(\_,\_,\_,\_)$ of the problem.  
\end{theorem}

First, we show that $\beta(\_,\_,\_,\_) \in$ \textit{NP}. To do so we give a polynomial time verification. Start by iterating through all objectives and creating a set of all objective cards. Then, iterate through the entire sequence of tricks, making sure all earlier definitions are followed. While iterating, mark off any objectives that are completed. Finally, check to make sure all objectives are satisfied. Because $l$ is bounded by $n$, this algorithm runs at worst in $\mathcal{O}(n^2)$, thus we can concluded $\beta(\_,\_,\_,\_) \in$ \textit{NP}.

\begin{lemma}
The Crew, $\beta(\_,\_,\_,\_)$, is NP-Hard.
\end{lemma}

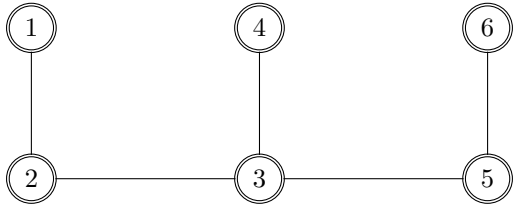
\begin{figure}
    \centering
    \begin{tikzpicture}[every circle node/.style={draw,double}]

\node[circle] (1) at (0,2) {1};
\node[circle] (2) at (0,0) {2};
\node[circle] (3) at (3,0) {3};
\node[circle] (4) at (3,2) {4};
\node[circle] (5) at (6,0) {5};
\node[circle] (6) at (6,2) {6};

\path[] (1) edge (2)
    (2) edge (3)
    (3) edge (4)
    (3) edge (5)
    (5) edge (6);

\end{tikzpicture}

    \caption{An instance of the Hamiltonian path problem on $G$, in which no Hamiltonian path exists.}
    \label{graph}
\end{figure}

To show that $\beta(\_,\_,\_,\_)$ is NP-Hard, we will use a proof by reduction. Our reduction is from the Hamiltonian path problem, where, given a undirected graph $G$, the problem is to find a whether a Hamiltonian path exists in $G$. The problem is known to be NP-Complete \cite{nphard}. To transform an instance of HP into $\beta(\_,\_,\_,\_)$, we use the following gadget. We  represent each vertex $v_i \in G$ as an individual player $i$. We then deal cards $(1, i), (2, i), ..., (deg(v_i), i)$ to each player whose corresponding node is adjacent to $v_i$. It is important to note that the ordering in which these cards are dealt does not matter. We then deal player $i$ the card $(deg(v_i) + 1, i)$, and give them the objective where $c$ is the card $(deg(v_i) + 1, i)$ we have just dealt to them. Finally, we need to even out the hand size of each player. To do so, we deal a number of "junk" cards to player $i$ equal to: $|V| - deg(v_i) + 1$. An example of this reduction can be seen in Figure 3, with the graph being drawn in Figure 2. This reduction can be done in polynomial time, as we process each vertex once. We define a "junk" card to be a card with an arbitrary value and that does not share a suit with any of the cards used in the earlier part of the reduction. Assuming the worst case, where we have a perfect graph $K_n$, for each vertex we, at most, must process every other vertex once. This allows the reduction to be done in $\mathcal{O}(n^2)$ time.

We will now show that a valid play sequence exists in our instance of The Crew, if and only if there is a Hamiltonian Path in $G$. First notice, that all players have the same amount of cards, which is also equal to $p$. Additionally, notice that each player has a single objective, $l = p$, which means only a single objective can be completed in one trick. Thus, in order to have a successful play sequence, one objective must be completed during every trick. In order for an objective to be completed, the suit of $c_o$ must be lead, otherwise playing $c_o$ will not result in it winning the trick and completing the objective. If we assume $G$ has a Hamiltonian path, say $(v_1, v_2, ..., v_n)$, then our constructed instance of The Crew, will also have a successful play sequence. We determine $p_1 = \tau$, and have $p_1$ complete their objective by leading with their objective card. All players will either follow suit or play a junk card, and as $p_1$ has the highest card of suit $1$ they will win the trick. Then, $p_1$ will lead the card with suit $2$ and allow $p_2$ to complete their objective. This sequence continues until all objectives are completed, thus a valid play sequence exists. However, if no Hamiltonian path exists, we will eventually reach a point where a player $i$ only has junk cards in his hand. By definition these junk cards do not share a suit with any other card, thus making it impossible to find a valid play sequence.     

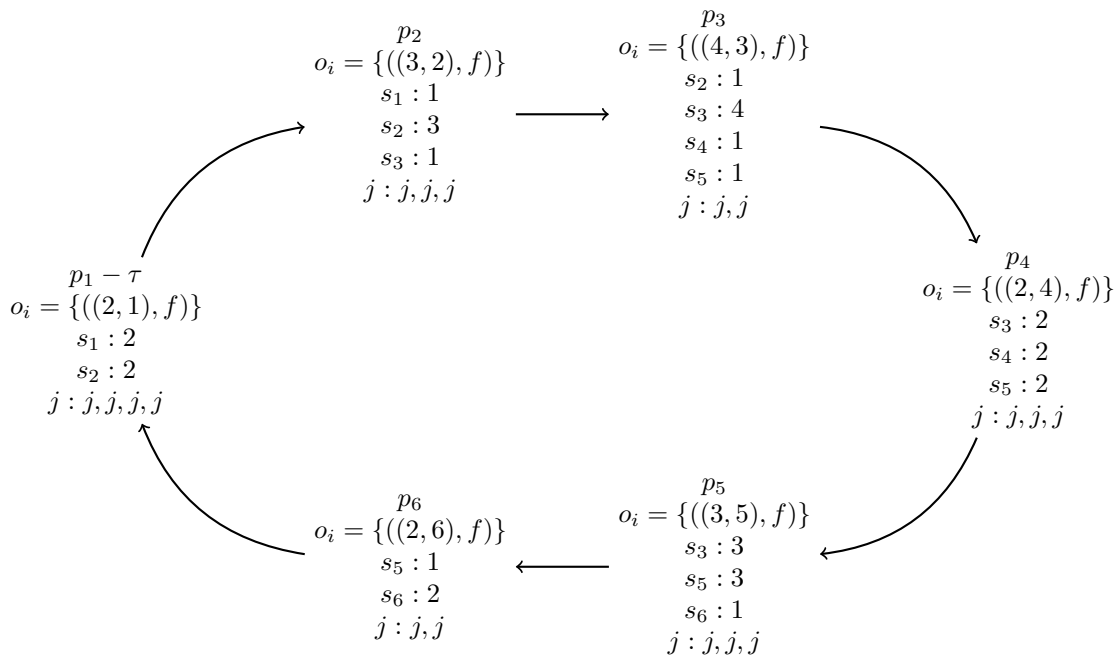
\begin{figure}
    \centering
    \begin{tikzpicture}

\node[align=center] (p1) at (0,0) {$p_1 - \tau$\\$o_i=\{((2,1),f)\}$\\$s_1:2$\\$s_2:2$\\$j:j,j,j,j$};
\node[align=center] (p2) at (4,3) {$p_2$\\$o_i=\{((3,2),f)\}$\\$s_1:1$\\$s_2:3$\\$s_3:1$\\$j:j,j,j$};
\node[align=center] (p3) at (8,3) {$p_3$\\$o_i=\{((4,3),f)\}$\\$s_2:1$\\$s_3:4$\\$s_4:1$\\$s_5:1$\\$j:j,j$};
\node[align=center] (p4) at (12,0) {$p_4$\\$o_i=\{((2,4),f)\}$\\$s_3:2$\\$s_4:2$\\$s_5:2$\\$j:j,j,j$};
\node[align=center] (p5) at (8,-3) {$p_5$\\$o_i=\{((3,5),f)\}$\\$s_3:3$\\$s_5:3$\\$s_6:1$\\$j:j,j,j$};
\node[align=center] (p6) at (4,-3) {$p_6$\\$o_i=\{((2,6),f)\}$\\$s_5:1$\\$s_6:2$\\$j:j,j$};

\path[->] (p1) edge [bend left, thick] (p2)
    (p2) edge [thick] (p3)
    (p3) edge [bend left, thick] (p4)
    (p4) edge [bend left, thick] (p5)
    (p5) edge [thick] (p6)
    (p6) edge [bend left, thick] (p1);

\end{tikzpicture}

    \caption{An instance of The Crew generated through the reduction from the graph in Figure 2.}
    \label{fig3}
\end{figure}

\begin{corollary}
The Crew, $\beta(\_,\_,\_, O*)$, is also NP-Hard.
\end{corollary}

Notice that in our reduction, all objective cards are owned by the player, thus we can bound our objective set more preciously and conclude that $\beta(\_,\_,\_, O*)$ is also NP-Hard.

\begin{corollary}
The Crew, $\beta(\_,\_,\_, \_)$, is NP-Hard even with the addition of a trump suit.
\end{corollary}

In section 2, we defined our model such that there was no trump card mechanic. Recall that trump cards can only be played when a player cannot follow suit $\forall c \in C_i, y_c \neq y_{c_\tau}$, and when played take priority over non trump cards. We chose not to use trump cards in our model for simpler analysis, however, with this particular reduction we can easily show that even with the addition of trump cards The Crew is still NP-Complete. To do so we alter the reduction slightly. After dealing junk cards, we will deal an arbitrary number of trump cards to all but a single player. We will now show that this does not change the validity of the reduction. As stated in corollary 6.3, our objective set can be bounded slightly to $O*$. Much like in section 4, this means that in order for an objective to completed, the objective card must win the trick. Therefore if a trump card is played, it is impossible for an objective to be completed in said trick. Additionally, the rules stated in our earlier proof still stand, moreover, that an objective must be completed every trick in order to have a successful play sequence, as one player still only has $p$ cards. Thus, if a trump card is played and no objective is completed it cannot be a valid play sequence. Finally, as players will always have the option to play a junk card when needed, adding trump cards does not stop a valid play sequence from existing. This completes the proof of corollary 6.4.  

\begin{theorem}
The unbounded version of The Crew is still NP-Complete even with the addition of task tokens.
\end{theorem}

As mentioned in the intro in The Crew: The Quest for Planet Nine, certain mission use task tokens which add another layer of complexity to the game. We will now show that the addition of these task tokens does not modify the complexity of the game. It is important to note that we make no assumptions about the rest of the objectives. 

Before proving that The Crew is still NP-Complete with task tokens, we must first define how these task tokens modify the game. In The Crew there are three different types of task tokens: tokens that require the first $i$ completed objectives to be specific objectives, a token that requires the objective to be completed last, and tokens that require an objective to be completed after a different objective. Note, that none of these objectives set requirements for tricks, only objectives.  For our purposes we will combine all such tokens into a single definition. Let a task token $t$ be a tuple $(b, a, o)$, where $b$ is the set of all objectives that must come before the mapped objective, $a$ is the set of all objectives that must come after the mapped objective and $o$ is the objective the task token is assigned to. Notice that all types of tokens found in the standard game can be created from the definition. For the first type of token we can use the following construction. Assign $b$ to $\emptyset$ and assign $a$ to be all other objectives. We can then continue this construction for the $i$th objective by assigning $b$ to all objectives already given a task token of this nature and $a$ to all other objectives. For the second type of task token, just assign $b$ to be all objectives, and for the third assign $b$ to be what ever objective needs to come before it.

With task tokens properly defined, we can now show that the addition of task tokens does not change the NP-Completeness of The Crew: The Quest for Planet Nine. To do so, we we will slightly modify the reduction used to prove lemma 6.2. After completing the reduction, we add a new player $i$, and deal them $p-1$ junk cards and the card $(i,i)$. We also give them an objective $c$  that maps to $(i,i)$, and has the task token with $b = O \setminus O_i$. Finally, we deal all other players $(f,i)$ where $f$ is some number $<p$, that has not been dealt already.

We will now show that this addition to reduction does not change the validity of the reduction. First, we show that a valid Hamiltonian Path implies a valid play sequence. For completing all but the the newest objective, we can use the same argument from lemma 6.2. As for completing objective $c$, all players have a card with suit $i$, thus as long as all other objectives have been completed, we can complete objective $c$. We will now show that this reduction does not cause false positives. First, notice that $p=l$, thus every trick must complete an objective. Second, notice that our objective set is still $O*$, meaning that only 1 objective can be completed, thus the same restrictions from earlier still apply. Finally, notice that if player $i$ takes any trick other than the last trick, the play sequence can't be valid. This is because player $i$ only has junk cards and $(p,i)$, neither of which can complete an objective. Therefore, the only way for a valid play sequence to exists is through the line of play described earlier. Thus, we can conclude that the addition of task tokens into The Crew: The Quest for Planet Nine does not effect its complexity upper bound.

\section{Conclusions and Future Work}

In this work, we have looked at the computational complexity of various generalized versions of The Crew: The Quest for Planet Nine. For the general unbounded version, we have shown the problem is NP-Complete. Additionally, various version of the problem can also be solved in polynomial time with respect to $n$, $p$, and $l$.   

However, several question regarding the complexity of the game remain unsolved, largely due to the problem being rich in parameters. A number of other versions of the problem could be studied, including a bounded number of players. A number of variations can also be created with the use of Task Tokens, including one in which all objectives must be completed in a specific order. Additionally, looking at the problem from a parameterized complexity may also provided interesting results. 

\printbibliography
\newpage
\appendix

\section{Rules of The Crew: The Quest for Planet Nine}

In this section we give the full rule set for The Crew: The Quest for Planet Nine. Access to the original rule set can also be found at \cite{bggCrew}

\subsection*{Components}

40 Large cards (playing cards)
\begin{itemize}
    \item 36 Color cards in four colors with values 1-9
    \item 4 Rocket cards with values 1-4
\end{itemize}
36 Small cards (objectives)
\begin{itemize}
    \item 36 Color cards in four colors with values 1-9
\end{itemize}
16 Tokens
\begin{itemize}
    \item 10 task tokens
    \item 5 radio communication tokens
    \item 1 distress signal token
\end{itemize}
1 commander token

\subsection*{Overview}

The Crew is a cooperative, mission-based trick-taking game. In it, all players work together to complete a shared goal. The specific goal is determined by the 1 of 50 missions included in the logbook. Many include drawing objectives from the small deck. If all players complete all objectives\footnote{The official rules use tasks instead of objectives. However, to match terminology used in the paper we will refer to them as objectives here.}, they win the mission. However, if a single objective is failed, or a player runs out of cards, all players lose.

\subsection*{Trick Taking}

As mentioned in the Overview, The Crew is a trick-taking game. In traditional trick taking games, the deck of cards is dealt to all players. Then, each player plays one of his/her/their own cards face up in the middle. The processes of each player playing a card one by one is referred to as the "trick". Once all players have played a single card, the player who places the card with the highest value wins the trick, as long as that card has "followed suit".

In The Crew, there are 5 different card suits: pink, blue, green yellow and rocket. In order to follow suit, a player must match the suit of the first player with the card they are playing. You must follow suit if you are able, however if you do not have a card of the lead suit, you may play a card from another suit. 

Unfortunately, doing so removes your ability to take the trick, unless you have played a rocket card. Rocket cards are "trump" cards. This means that, when played, the rocket suit takes precedent, and now the highest rocket card played in the trick wins the trick. During a trick, a player is under requirement to play a specific card, as long as the follow suit when necessary. This means that a rocket card can be played when a color suit was lead, and still take the trick. If a rocket card is led, then that becomes the led suit, and players must follow if possible.

After determining who has won the trick, give all cards to the player who has won the trick. That player then sets those cards face down, and players can only look at the cards from the most recently completed trick.

\subsection*{Communication}

One of the critical mechanics in The Crew is the limited communication between players. During play, players are not allow to give any information about what cards they posses. The only way to communicate this information is through radio communication tokens. During set up, each player is dealt one token, to be used exactly once during the mission. Communication tokens can be used before a trick, but not during one. To use the token, a player picks one of the color (non-rocket) cards from their hand and places it face up. The card is still a part of the players hand, and can be played as normal. Then place the radio communication token on the card in one of three positions to convey the following information.
\begin{itemize}
    \item At the top, if this your highest card of the color
    \item In the middle, if this the only card you have of the color
    \item At the bottom, if this is your lowest card of the color.
\end{itemize}

It is important to note that one of these three conditions must apply, otherwise you are not allowed to communicate with that card. You may also not communicate with rocket cards. Once placed the token may not be changed even if the statement becomes incorrect. For example, your lowest card may become the only card during play.

\subsection*{Set Up}
Before play, players perform the following actions in order to set up the game.
\begin{enumerate}
    \item Shuffle the 40 card playing deck and deal them as equally as possible
    \item Each person takes a radio communication token
    \item Place the distress signal token face down
    \item Shuffle the 36 card objective deck and keep them face down.
\end{enumerate}
Now players can set up the mission they have elected to play. Most missions will contain a certain number of objectives, indicated by the task book. To set up the objectives, draw the specified number of objective cards and place them face up on the table. Next, the player with the four rocket announces they have the card, and is assigned the commander for the current mission. Then players take turns in play (clockwise) order selecting an objective and placing it front of themselves. 

\subsection*{Completing Objectives \& Playing the Game}
With all set up completed, players can now begin the game, with the commander leading the first trick. For all future tricks, the player who won the previous trick leads. During play, players will try and complete objectives. To complete an objective, a player must win a trick containing the playing card that is matched on of their objectives. It is important to note that a player can fulfill several objectives with the same trick. Once all objectives are completed the players have won! However, if a player wins a single playing card for which another player has a corresponding objective, you lose immediately. 

\subsection*{Mission Modifiers}
To keep gameplay interesting, The Crew introduces a number of smaller mechanics that are used in specific missions to provide variation in play. Here we will give an overview of all the ones discussed in the rule book. Note that there are other "mechanics" used in the missions than just the ones defined here, however these are typically only used in a single mission.
\begin{itemize}
    \item Task Tokens - In many of the missions, task tokens will be used to add variety and challenge. Each token is assigned to a single task card and add other conditions for completing the objective. There are 3 types of task tokens. The first, require that a task be complete first, second, third etc. It should be noted, however, that these are only used in front of other tasks, never after. The second, requires that a task be completed last. The final type requires that a task be completed after another task.
    \item Dead Zone - Missions with the dead zone symbol reduce communication between the players. Players can now no longer specify which condition they have satisfied to communicate a card.
    \item Disruption -If a missions has the disruption symbol, players are not allowed to communicate for the first few tricks. The specific number is listed on the mission and varies mission to mission.
    \item Commanders Decision - When a mission has this modifier, all tasks are dealt face down. Each player then answers if the feel they can take on all of the tasks. Answers may only be "yes" or "no". The commander than chooses a player and gives them all dealt objectives. 
    \item Commanders Distribution - In this instance, the commander chooses how to distribute the objective cards. To do so, they deal the objective tokens face down. Then one by one, reveal an objective and ask each player if they want to take the objective. Again, they may only answer "yes" or "no". Then the commander chooses who receives the task token. Finally, the commander must also distribute evenly, as at the need of distribution no player can have 2 or more tasks than another player.  
\end{itemize}
\end{document}